# Magnetize Topological Surface States of $Bi_2Se_3$ with a $CrI_3$ Monolayer

Y. S. Hou, and R. Q. Wu

Department of Physics and Astronomy, University of California, Irvine, CA 92697-4575, USA

## Abstract

To magnetize surfaces of topological insulators without damaging their topological feature is a crucial step for the realization of the quantum anomalous Hall effect (QAHE), and still remains as a challenging task. Through density functional calculations, we found that adsorption of a two-dimensional van der Waals (2D-vdW) ferromagnetic insulator $CrI_3$ monolayer can create a sizable spin splitting at the Dirac point of the topological surface states of $Bi_2Se_3$ films. Furthermore, general rules that connect different quantum and topological parameters are established through model analyses. This work provides a useful guideline for the realization of QAHE at high temperature in heterostructures of 2D-vdW magnetic monolayers and topological insulators.

Email: wur@uci.edu

Topological insulators (TIs) are emergent quantum materials that have nontrivial band gaps in their bulks along with topologically protected surface or edge states at boundaries [1, 2]. For the design of the next-generation spintronic devices [3-6], especially those based on the quantum anomalous Hall effect (QAHE), it is crucial to find efficient ways to magnetize their topological surface states (TSSs) and yet maintain their topological features [7-9]. The conventional approach is doping magnetic ions into TIs [10-22], and the QAHE has been successfully realized in Cr- or V-doped (Bi, Sb)$_2$Te$_3$ thin films [20-22]. However, it is still very challenging to control the distribution and magnetic order of dopants in TIs, and the critical temperature for the observation of QAHE is extremely low (30 mK) [20]. A promising alternative way to magnetize TSSs is through the interfacial magnetic proximity effect by putting three-dimensional (3D) magnetic insulators on TIs [23-34]. Unfortunately, the interfacial hybridization in most of such heterostructures is too strong and the TSSs are either damaged or shifted away from the Fermi level [27, 29, 35]. To this end, recently discovered two-dimensional (2D) van der Waals (vdW) ferromagnetic monolayers (MLs) such as CrI$_3$ [36] appear to offer an optimal way to magnetize the TSSs of TIs. First, the Curie temperature of the CrI$_3$ ML is as high as 45 K [36] and it is hence conceivable that the critical temperature of the QAHE in CrI$_3$/TI heterostructures can be far higher than that in the Cr-doped (Bi, Sb)$_2$Te$_3$ thin films [20]. Furthermore, the CrI$_3$ ML is a semiconductor and thus the transport properties of TIs should be preserved. Nevertheless, magnetic ions in most 2D-vdW magnetic materials such as Cr$^{3+}$ ions in CrI$_3$ are typically covered by non-magnetic layers in both sides; it is thus questionable if their spin polarization can be sensed by the TSSs, even though the latter have fairly large spatial extension.

In this Letter, we report results of systematic computational studies based on the density functional theory (DFT) that suggest the possibility of utilizing CrI$_3$ ML to magnetize the TSSs of a prototypical 3D TI: Bi$_2$Se$_3$ (BS). We build up CrI$_3$/BS/CrI$_3$ heterostructures with a varying thickness for the BS films, from three to seven quintuple layers (QLs). We find that the TSSs of BS can be effectively magnetized by the CrI$_3$ ML in all cases. However, we show that CrI$_3$/BS/CrI$_3$ becomes a Chern insulator only when the BS film is thicker than five QLs. This results from the competition between the exchange field from CrI$_3$ and the remaining interaction between two surfaces of BS. This work reveals the subtleness of designing topological spintronic materials and provides a general guidance for the realization of the QAHE at high temperature by combining 3D TIs with 2D vdW magnetic MLs [36, 37].

We use the Vienna *Ab Initio* Simulation Package at the level of the generalized gradient approximation [38-41] in this work. The projector-augmented wave pseudopotentials are adopted to describe the core-valence interaction [42, 43] and the energy cutoff for the plane-wave expansion is set to 500 eV [41]. As shown in Fig. 1a, we construct inversion symmetric CrI$_3$/BS/CrI$_3$ slab models with a $\sqrt{3}\times\sqrt{3}$ BS supercell in the lateral plane to match the CrI$_3$ ML. The vacuum space between adjacent slabs is set to 15 Å. Atomic structures are fully optimized with a criterion that requires the force on each atom being less than 0.01 eV/Å. To correctly describe the weak interaction across CrI$_3$ and BS layers and the strong relativistic effect in Bi and I atoms, we include the nonlocal vdW functional (optB86b-vdW) [44, 45] and the spin-orbit coupling (SOC) term in the self-consistent iterations. Furthermore, the LSDA+U method [46], with the on-site Coulomb interaction U=3.0 eV and the exchange interaction J=0.9 eV, is adopted to take the strong correlation effect of Cr 3*d* electrons into account.

CrI$_3$ ML is a robust 2D insulating ferromagnet. It has a perpendicular magnetic anisotropy and a reasonably high Curie temperature (45 K) [36]. Structurally, CrI$_6$ octahedrons form a honeycomb lattice. The optimized lattice constant of the pristine CrI$_3$ ML is $a_{\text{CrI}_3-\text{ML}}$ = 7.04 Å, consistent with the previous theoretical result [47]. This value is only 2.5% smaller than the size of the $\sqrt{3}\times\sqrt{3}$ BS supercell (7.22 Å) and hence we stretch the CrI$_3$ ML so as to use a manageable unit cell for the simulation of CrI$_3$/BS/CrI$_3$. We find that all main properties of CrI$_3$ ML are not significantly affected by the small lattice stretch [see Part I in Supplementary Material (SM)].

There are three possible highly symmetric alignments between the $\sqrt{3}\times\sqrt{3}$ BS supercell and the CrI$_3$ ML, namely, with Cr ions taking the Se, hollow, or Bi sites on BS (Fig. 1a), respectively. The calculated binding energies suggest that $Cr^{3+}$ cations prefer to sit on the top of $Se^{2-}$ anions (see Part II in SM). The optimized interlayer distances between the $CrI_3$ ML and the BS surface are in a range from 3.07 to 3.12 Å, depending on the thickness of the BS film (see Part II in SM). Hereafter, we only discuss electronic and magnetic properties of the most stable configuration.

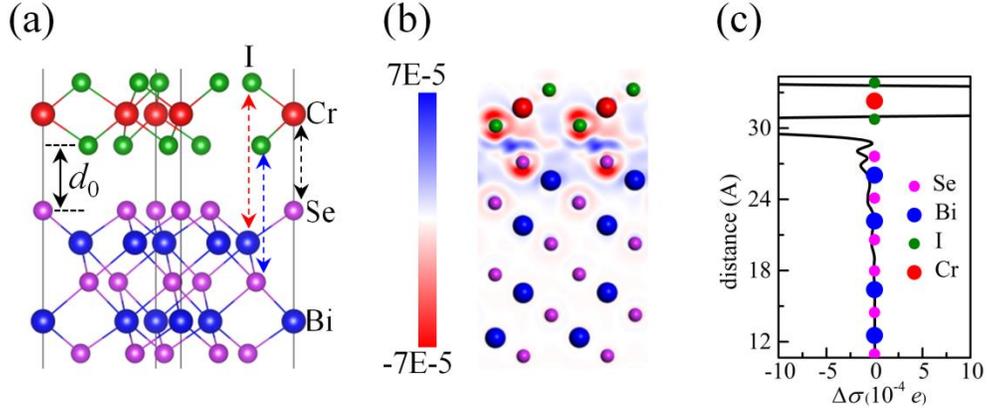

*Figure 1 (Color online) (a) Side view of the most stable $CrI_3$/BS/$CrI_3$ heterostructure. The Green, red, pink and blue balls are for I, Cr, Se and Bi atoms, respectively. Dashed-lines show the alignments between atoms in $CrI_3$ and $Bi_2Se_3$ layers; $d_0$ denotes the optimized vdW gap. (b) Real-space distribution of the charge difference $\Delta\rho = \rho_{total} - \rho_{BS} - \rho_{CrI_3}$ and (c) planar-averaged spin density $\Delta\sigma = \rho_\uparrow - \rho_\downarrow$ in the interfacial region of heterostructure $CrI_3$/6QL-BS/$CrI_3$.*

To shed some light on the interaction across the $CrI_3$/BS interface, we plot the charge density difference $\Delta\rho$ of $CrI_3$/BS/$CrI_3$, using 6QL-BS as an example in Fig. 1b. The charge redistribution only occurs within a couple atomic layers in BS and the magnitude of $\Delta\rho$ is small. In particular, there is no observable net charge transfer between $CrI_3$ and BS. This is understandable since the band gap of the freestanding $CrI_3$ ML is rather wide (larger than 1 eV). The interfacial Se atoms acquire a small but meaningful magnetic moment of -0.003 $\mu_B$, which aligns antiparallelly with the magnetic moments of $Cr^{3+}$ ions. The planar-averaged spin density $\Delta\sigma$ (Fig. 1c) shows that the negative spin polarization penetrates through the top half of the first QL of BS, following the direction of spin polarization of the iodine atoms.

The calculated band structures of $CrI_3$/BS/$CrI_3$ with different thicknesses of BS (four, five and six QLs) are shown in Fig. 2a, 2b and 2c, respectively. One important feature is that bands near the Fermi level are predominantly from BS and $CrI_3$ states lie either 0.7 eV above the Fermi level or at least 0.3 eV below the Fermi level. As a result, the Dirac cone feature of TSSs are well maintained, or more explicitly (I) all three cases have band gaps of several meV at the $\Gamma$ point; (II) the spin degeneracies of the TSS are lifted, indicating that the TSSs of BS are magnetized by the $CrI_3$ ML. These results suggest the suitableness of using vdW magnetic MLs to realize QAHE, instead of using conventional ferromagnetic or antiferromagnetic films that damage the TSSs [29, 30, 35, 48].

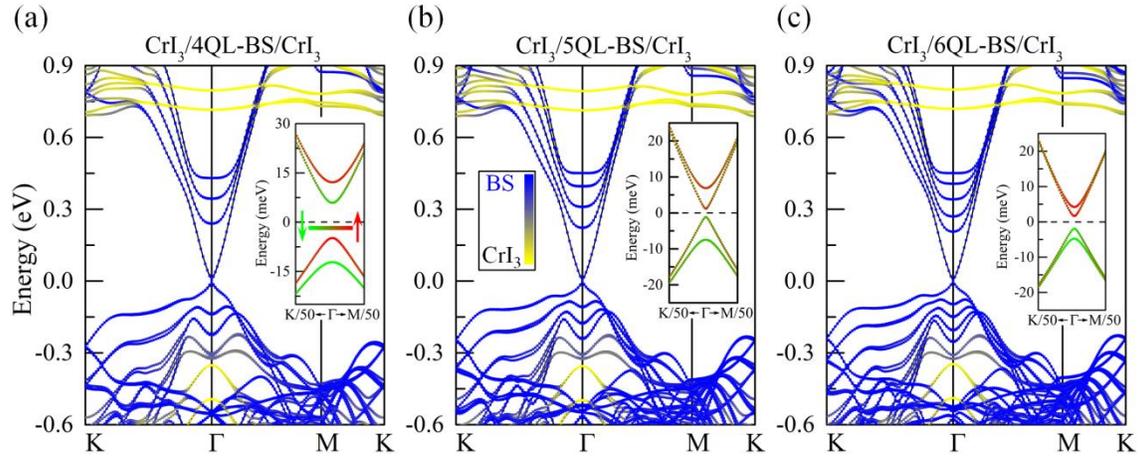

*Figure 2. (Color online) DFT calculated band structures of (a) $CrI_3$/4QL-BS/$CrI_3$, (b) $CrI_3$/5QL-BS/$CrI_3$, (c) $CrI_3$/6QL-BS/$CrI_3$. Insets in (a), (b) and (c) are the fine band structures around the Fermi level. Colors in the main panels indicate the weights of bands from BS (blue) and $CrI_3$ (yellow). Colors in insets indicate the spin projections.*

To thoroughly study the topological properties of $CrI_3$/BS/$CrI_3$, one needs to directly calculate their Berry curvatures and Chern numbers. With the sizes of unit cells in the present work, these calculations are extremely time-demanding even using the Wannier90 package. We note that states of $CrI_3$ ML are far away from the Fermi level and the effect of $CrI_3$ ML on TSSs of BS can be represented by an exchange field ($H_{Zeeman}$) and an interfacial potential ($H_{Interface}$). Therefore, the topological properties of TSSs can be described by an effective four-band Hamiltonian [3, 8, 9, 49] with a basis set of $\{|t,\uparrow\rangle, |t,\downarrow\rangle, |b,\uparrow\rangle$ and $|b,\downarrow\rangle\}$ as:

$$H(k_x,k_y) = H_{surf}(k_x,k_y) + H_{Zeeman}(k_x,k_y) + H_{Interface}(k_x,k_y)$$

$$= A(k_x^2 + k_y^2) + \begin{bmatrix} 0 & iv_F k_- & M_k & 0 \\ -iv_F k_+ & 0 & 0 & M_k \\ M_k & 0 & 0 & -iv_F k_- \\ 0 & M_k & iv_F k_- & 0 \end{bmatrix} + \begin{bmatrix} \Delta & 0 & 0 & 0 \\ 0 & -\Delta & 0 & 0 \\ 0 & 0 & \Delta & 0 \\ 0 & 0 & 0 & -\Delta \end{bmatrix} + \begin{bmatrix} V_p & 0 & 0 & 0 \\ 0 & V_p & 0 & 0 \\ 0 & 0 & -V_p & 0 \\ 0 & 0 & 0 & -V_p \end{bmatrix} \quad (1)$$

Here, $t$, $b$ denote the top and bottom surface states and $\uparrow$, $\downarrow$ represent the spin up and down states; $v_F$ is the Fermi velocity; $k_\pm = k_x \pm ik_y$ are wave vectors; $M_k$ is a finite mass term [$M_k = M - B(k_x^2 + k_y^2)$] that describes the interaction between surfaces of BS; $\Delta$ is the exchange field from CrI$_3$; and $V_p$ represents the magnitude of the asymmetric interfacial potential. Because inversion symmetry is preserved in our CrI$_3$/BS/CrI$_3$, $V_p$ is vanishing. In experiments, if inversion symmetry breaks due to substrates, $V_p$ will be nonvanishing and its effect is discussed in details below. As shown in Part III of SM, this Hamiltonian fits very well the DFT bands of CrI$_3$/BS/CrI$_3$ around the $\Gamma$ point, indicating its applicability to these systems.

The Berry curvature $\Omega(\mathbf{k})$ and Chern number $C_N$ are calculated based on the following formulas [50, 51]:

$$\Omega(\mathbf{k}) = \sum_n f_n(\mathbf{k})\Omega_n(\mathbf{k}) \quad (2),$$

$$\Omega_n(\mathbf{k}) = -2\,\mathrm{Im}\sum_{m\neq n} \frac{\langle\psi_{n\mathbf{k}}|v_x|\psi_{m\mathbf{k}}\rangle\langle\psi_{m\mathbf{k}}|v_y|\psi_{n\mathbf{k}}\rangle\hbar^2}{(E_{m\mathbf{k}} - E_{n\mathbf{k}})^2} \quad (3),$$

$$C_N = \frac{1}{2}\sum_n \int_{BZ} \Omega_n(\mathbf{k})d^2\mathbf{k} \quad (4).$$

The summation in Eq. (2) goes over all of the occupied states. $f_n$ is Fermi-Dirac distribution function, and $E_{n\mathbf{k}}$ is the eigenvalue of the Bloch functions $|\psi_{n\mathbf{k}}\rangle$, and $v_{x(y)}$ are the velocity operators. The QAHE can be observed in materials that have $C_N \neq 0$.

It is interesting to see that there is a clear phase boundary in Fig. 3a for the topological feature of $CrI_3/BS/CrI_3$, i.e., $C_N=0$ ($C_N=1$) when BS is thinner (thicker) than five QLs. That is to say, $CrI_3/BS/CrI_3$ heterostructures with less than five QLs of BS are normal insulators and five QLs of BS are required to realize the QAHE in experiments. The calculated band gaps of $CrI_3/5QL$-$BS/CrI_3$, $CrI_3/6QL$-$BS/CrI_3$ and $CrI_3/7QL$-$BS/CrI_3$ are 2.3 meV (27 K), 3.6 meV (42 K) and 5.9 meV (68 K), respectively. Considering that the ML $CrI_3$ has a $T_C$ of 45 K [36] and no other complex factors, such as uncontrollable distribution of dopants, local magnetic ordering and charge transfer, are involved in these systems, we believe that QAHE should be observable in $CrI_3/BS/CrI_3$ heterostructures up to a few tens Kelvins, much higher than the temperature achieved with the doping approach [20-22].

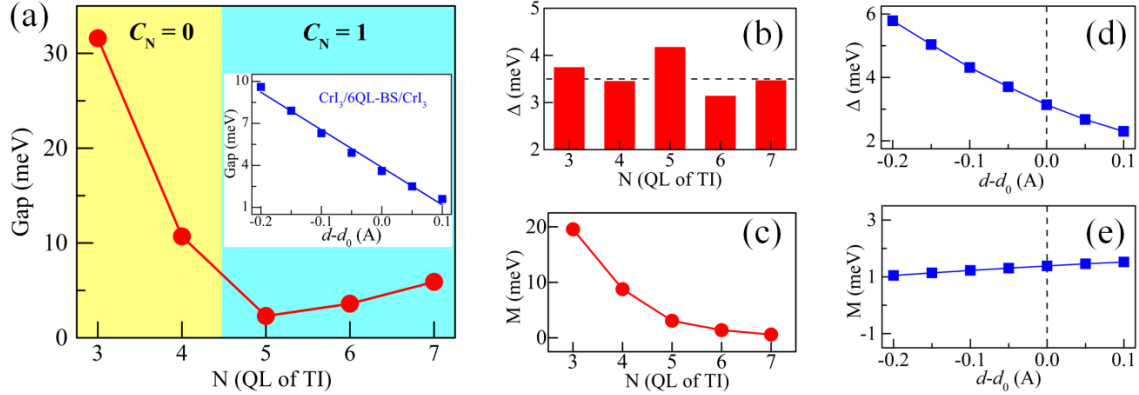

*Figure 3 (Color online) (a) Dependence of Chern numbers and gaps of $CrI_3/BS/CrI_3$ on the number (N) of QLs of BS. Inset shows the dependence of band gaps of $CrI_3/6QL$-$BS/CrI_3$ on the $CrI_3$-BS interlayer distance d. (b) and (c) show the fitting parameters $\Delta$ and M in different $CrI_3/BS/CrI_3$ heterostructures, respectively. The horizontal dashed line in (b) shows the position of 3.5 meV. Dependences of $\Delta$ and M of $CrI_3/6QL$-$BS/CrI_3$ on d are shown in (d) and (e), respectively. Vertical dashed lines in (d) and (e) show the optimized $d_0$.*

In general, we find rules that govern topological properties of $CrI_3/BS/CrI_3$, i.e., (1) $C_N=1$ when $\Delta^2 > M^2$ and (2) $C_N=0$ when $\Delta^2 < M^2$. This is consistent with the phase diagram in Ref. [8] and gives simple ways to manipulate the topological properties of systems with a 2D magnetic monolayer on TI. As shown in the Fig. 3b and 3c, the exchange filed $\Delta$ slightly depends on the thickness of BS and fluctuates around 3.5 meV. This result is understandable since exchange filed $\Delta$ mainly depend on the interfacial interaction between CrI$_3$ and BS. As expected, M has a strong dependence on N and exponentially decreases as the BS film becomes thicker. Therefore, the topological phase transition is mainly cause by the remaining interaction of the two TSSs across the BS film.

Considering that the interfacial distance, *d*, in experiments may differ from our DFT result, we also examine the robustness of the topological property by changing *d* in a small range for CrI$_3$/6QL-BS/CrI$_3$. As shown in the Fig. 3d and 3e, we find that only $\Delta$ sensitively change with *d*, whereas M hardly change. Importantly, $\Delta^2 > M^2$ holds in the range from $d_0$-0.2 to $d_0$+0.1 Å, where $d_0$ denotes the DFT optimized interfacial distance. Note that the band gaps of CrI$_3$/6QL-BS/CrI$_3$ noticeably increase with reducing the vdW gap (see the inset in Fig. 3a). Therefore, CrI$_3$/6QL-BS/CrI$_3$ should easily manifest QAHE according to our calculations.

As mentioned above, the inversion symmetry of CrI$_3$/BS/CrI$_3$ can be broken by substrates in experiments. To explore the effect of inversion symmetry breaking on the topological properties of CrI$_3$/BS/CrI$_3$ heterostructures, we add a nonvanishing $V_p$ in the effective four-band Hamiltonian and take CrI$_3$/6QL-BS/CrI$_3$ as an example. With increasing $V_p$ from zero to 5.0 meV, the original topologically nontrivial band (Fig. 4a) involves into a M- and W-shaped band (Fig. 4c) through a semimetal band at $V_p$ = 2.8 meV (Fig. 4b). Surprisingly, this M- and W-shaped band is a topologically trivial insulator although band inversion is usually suggested in such band. The bottom panel of Fig. 4d shows the occupied bands Berry curvature of $V_p$=5.0 meV around the Γ point. One can observe that not only huge negative Berry curvature but also large positive Berry curvature appears. This is completely different from the Berry curvature of $V_p$=0 in which only huge positive Berry curvature appears around the Γ point (see the upper panel of Fig. 4d). By integrating the Berry curvature in the first Brillouin zone for $V_p$=5.0 meV, we obtain that the total Berry curvature is cancelled out so that Chern number $C_N$ is zero. So the M- and W-shaped band with $V_p$=5.0 meV is a norm insulator according to the criterion discussed in the previous paragraph. This shows the subtleness of identifying a topological state of CrI$_3$/BS/CrI$_3$ heterostructures, or probably all analogous material systems. Based on a systematical study (Fig. 4e), we unveil that a topological phase transition from Chern insulator to norm insulator takes place at the critical point [8] $\Delta^2 = M^2 + V_p^2$: (A) when $\Delta^2 > M^2 + V_p^2$, $C_N$=1; (B) when $\Delta^2 < M^2 + V_p^2$, $C_N$=0. It is worth noting that introduction of $V_p$ reduces the topologically nontrivial gap in the Chern insulator region (Fig. 4e). Therefore preservation of the inversion symmetry is of importance to realize the high-temperature QAHE in CrI$_3$/BS/CrI$_3$ heterostructures in experiments.

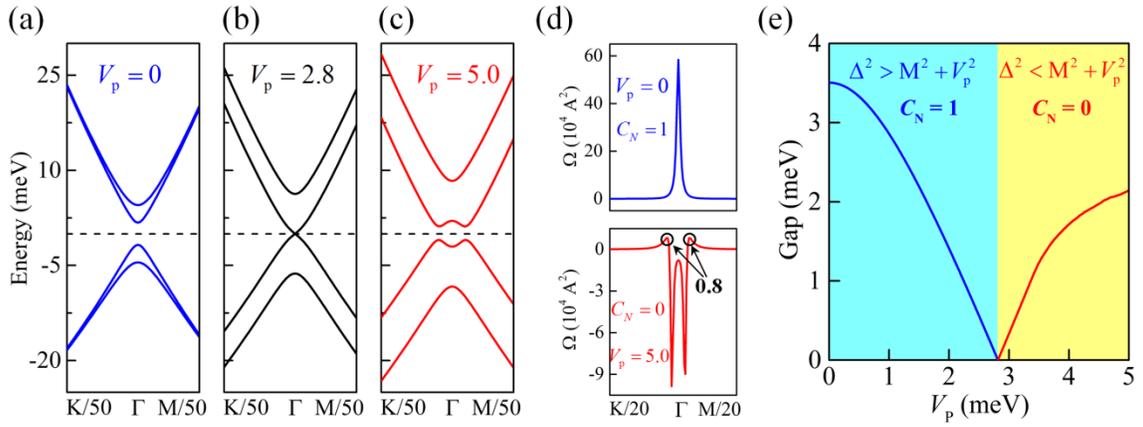

*Figure 4 (Color online) Effect of $V_p$ on the topological properties of CrI$_3$/6QL-BS/CrI$_3$. (a)-(c) show band evolutions with $V_p$ increasing from zero to 5.0 meV. (d) Occupied bands Berry curvatures $\Omega$ around the $\Gamma$ point for $V_p$=0 (upper panel, corresponding to (a)) and $V_p$=5.0 meV (bottom panel, corresponding to (c)). The large positive Berry curvatures in the bottom panel are highlighted by the black circles. (e) Dependence of Chern numbers and gaps on $V_p$. $\Delta$ =3.1 meV and M=1.4 meV are adopted in (a)-(e).*

Undoubtedly, 2D ferromagnetism in vdW materials such as CrI$_3$ is entirely new [52, 53]. Our work clearly demonstrates that one of its outstanding applications is magnetizing TSSs of TIs without any detrimental effect on the Dirac cone of TIs. This is dramatically distinct from the conventionally adopted 3D magnetic insulators such as MnSe and EuS in previous magnetic insulators/TI heterostructures since they either damage TSSs or shift TSSs away from the Fermi level [27, 29, 35]. We also note that large spin splitting may be produced in TSSs through δ-codoping of Mn and Se(Te) in the topmost QL of TIs [54, 55]. An important merit of our systems here is that both $CrI_3$ and BS have stable vdW-type structures and there is no issue regarding doping, diffusion and magnetic ordering, which are inherent in doping approaches. Magnetizing TSSs of TIs is of great importance as it gives rise to many novel phenomena, especially the hallmark of 3D TIs—topological magneto-electrical effect [56].

In summary, using $CrI_3$/BS/$CrI_3$ heterostructures as modeling systems, we demonstrated that the emergent 2D vdW magnetic MLs can effectively magnetize TSSs of TIs and, importantly, also maintain their topological characteristics around the Fermi level. Furthermore, our analyses with an effective four-band Hamiltonian verify that $CrI_3$/BS/$CrI_3$ are Chern insulators when the BS film is six QLs or thicker. Their band gaps are a few meV, and we believe that the QAHE is observable in these heterostructures at a temperature of a few tens K. Obviously, even higher temperature QHAE can be achieved with other 2D vdW magnetic monolayers with stronger magnetization or larger spatial extension of spin density.

Work was supported by DOE-BES (Grant No. DE-FG02-05ER46237). Density functional theory calculations were performed on parallel computers at NERSC supercomputer centers.